\documentstyle{mnrasl}
\begin{document}
\newcommand{\TEFF}{\mbox{T$_{eff}$}}
\newcommand{\TSTAR}{\mbox{T$_{\ast}$}}
\newcommand{\TE}{\mbox{T$_e$}}
\newcommand{\NE}{\mbox{n$_e$}}
\newcommand{\HII}{\mbox{H~II}}
\newcommand{\etal}{\mbox{et al.}}
\newcommand{\MSUN}{\mbox{M$_{\odot}$}}
\newcommand{\Halpha}{\mbox{H$_{\alpha}$}}
\newcommand{\OII}{\mbox{[OII]}}
\newcommand{\OIII}{\mbox{[OIII]}}
\newcommand{\SII}{\mbox{[SII]}}
\newcommand{\NII}{\mbox{[NII]}}
\newcommand{\Mgtw}{\mbox{Mg$_2$~}}
\newcommand{\Hbeta}{\mbox{H$\beta$~}}
\newcommand{\Fetw}{\mbox{Fe52~}}
\newcommand{\Feth}{\mbox{Fe53~}}
\newcommand{\Lsun}{\mbox{L$_{\odot}$}}
\newcommand{\Zsun}{\mbox{Z$_{\odot}$}}
\newcommand{\Msun}{\mbox{M$_{\odot}$}}

\title[Nuclear Indices in Ellipticals]
{On the metallicity distribution in the nuclei of elliptical galaxies}

\author[L.~Greggio]
{L. Greggio $^{1,2}$\\
$^1$ Universitaets Sternwarte--Muenchen, Scheiner Str. 1, D-81679 Muenchen, 
FRG\\
$^2$ Dipartimento di Astronomia,Universit\'a di Bologna, I40100 Bologna, 
Italy}

\maketitle

\begin{abstract}
\leftmargin 5cm
Using current models of spectrophotometric properties of single age,
single metallicity stellar populations I have computed the  
\Mgtw, \Hbeta, \Fetw and \Feth line
strengths for stellar populations with a metallicity spread.
The comparison of these models with the nuclear indices
of early type galaxies yield the following major conclusions. The metallicity
distribution of the closed box, simple model for the chemical evolution of 
galaxies is not able to account for \Mgtw and \Fetw, \Feth values in excess 
of $\sim$ 0.27,3 and 2.7, respectively, which are 
observed in the nuclei of a large fraction of Ellipticals. 
To reproduce the line strengths in these galaxies high average metallicities, 
small metallicity dispersion and old ages are required. In particular,
\Mgtw values of $\sim$0.3 are reproduced only with a metallicity distribution
ranging from $\sim$ 0.5\Zsun to $\sim$ 3\Zsun, and 
$\sim$ 15 Gyr old stellar populations.
I interpret the data as indicating 
that the gas out of which the nuclei of ellipticals formed was pre-enriched, 
to larger metallicities for increasing \Mgtw.
The presence of a metallicity
dispersion does not alter the relation between \Mgtw and Iron
indices with respect to the SSP models.
Thus, the need for a Mg/Fe overabundance in the strongest lined galaxies
is confirmed, and I present a simple way to estimate the [Mg/Fe] ratio
on the basis of existing models with solar abundance ratios.

\end{abstract}

 
\section{Introduction}

The question of the age of the bulk of the stars in elliptical galaxies
is still subject to a large debate. Opposed to the classical picture in 
which Ellipticals are basically inhabited by $\sim$ 15 Gyr old stellar 
populations (see e. g. Renzini 1986), O'Connell (1986), amongst others, 
proposed that a substantial component of stars as young as 5 Gyr has to 
be present to account for the observed spectral energy distribution of
Ellipticals. In more recent years many observational evidences have been 
found to support the notion of elliptical galaxies formation at high redshift, 
including the tightness of the
colour$-$central velocity dispersion ($\sigma$) relation found for Ellipticals
in Virgo and Coma by Bower, Lucey $\&$ Ellis (1992) ; the thinness of the
fundamental plane (Renzini $\&$ Ciotti 1993) for the Ellipticals in the
same two clusters; the modest passive evolution measured for cluster
Ellipticals at intermediate redshifts (Franx $\&$ van Dokkum 1996, 
Bender, Ziegler $\&$ Bruzual 1996); 
the negligible luminosity evolution observed for the
red galaxies  (Lilly et al. 1995) and for early
type galaxies in clusters in the redshift range z $<$ 1 (Dickinson 1996); 
the detection
of bright red galaxies at redshifts as large as 1.2 (Dickinson 1996). 
On the other hand, hints for a continuous formation of Ellipticals
in a wide redshift range have also been found: the relatively 
large \Hbeta values measured in a sample of nearby Ellipticals, which
could indicate a prolongued star formation activity
in these galaxies, up to $\sim$ 2 Gyr ago (Gonzalez 1993, Faber et al. 1995);
the apparent paucity of high luminosity Ellipticals at z$\simeq$ 1 compared
to now (Kauffmann, Charlot $\&$ White 1996).
\par
Two competing scenarios have been proposed also for the process leading to
the formation of the bulk of the stars in 
Ellipticals: early merging of lumps containing gas and stars 
(e.g. Bender, Burstein $\&$ Faber 1993), in which some dissipation
plays a role in establishing the chemical structure of the outcoming
galaxy; and the merging of 
early formed stellar systems, occurring in a wide redshift range, and
preferentially at late epochs, following the
hierarchical formation of structures (Kauffmann, White $\&$ Guiderdoni 1993).
\par
In order to help understanding when and how elliptical galaxies formed I have 
computed synthetic spectral indices for stellar populations with a metallicity
spread, and compared them to the corresponding observations in the nuclei
of Ellipticals. 
The study of line strengths in the spectra of early type galaxies has
shown to be a powerful tool for investigating the age and the metallicity 
of these systems (Faber et al. 1995; Fisher, Franx and Illingworth 1995; 
Buzzoni 1995b and references therein). 
With few exceptions (Vazdekis et al. 1996, Bressan, Chiosi $\&$ Tantalo 1996), 
most of
the authors have interpreted the observed line strengths through comparisons
with theoretical models constructed for single age, single 
metallicity stellar populations (SSPs). The major results of these
studies can be sumarized as follows:
\par\noindent
i) the \Mgtw indices measured in elliptical galaxies are consitent with
the notion that these systems are inhabited by old stellar populations,
the differences in \Mgtw tracing differences in average metallicity
(Buzzoni, Gariboldi $\&$ Mantegazza 1992). However, the difficulty in 
determining separately age and metallicity (Renzini 1986) weakens
considerably this simple picture (Worthey 1994);
\par\noindent
ii) the \Hbeta line strength offers an opportunity to break the
age$-$metallicity degeneracy, if this index is 
measuring the temperature of turn-off stars (see Faber et al. 1995).
In this view, the data derived for a sample of nearby Elliptical galaxies 
indicate that the ages of the stellar populations in their nuclei
span a wide range, the weakest \Mgtw galaxies being the youngest 
(Gonzalez 1993, Faber et al. 1995);
\par\noindent
iii) the Magnesium to Iron abundance ratio in the nuclei of the highest
\Mgtw Ellipticals 
is likely to be larger than solar (Gorgas, Efstathiou $\&$ Arag\'on Salamanca
1990; Worthey, Faber $\&$ Gonzalez (1992), hereinafter WFG; Davies, Sadler 
$\&$ Peletier 1993).
Weiss, Peletier and Matteucci (1995) estimate [Mg/Fe] ranging from 0.3
to 0.7 dex within the brightest ellipticals.
\par
Real galaxies host composite stellar 
populations, with a spread in the major parameters like age and metallicity.
This may apply also to the nuclei of galaxies, where typically
$\sim$10$^7$ \Lsun are sampled, corresponding to $\sim$100 bright globular
clusters. 
The existence of a substantial metallicity spread in elliptical
galaxies and bulges is supported by direct observations. For example, 
the Colour-Magnitude
diagram of a field in M32 (Freedman 1989) shows a wide red
giant branch, corresponding to stars spanning a metallicity range of
0.6 dex approximately. The K-giants in the galactic bulge
have metallicities ranging from $\sim$ 0.1\Zsun to  
$\sim$ 5\Zsun (Rich 1988).
Finally, the mere evidence for abundance gradients in elliptical galaxies,
as inferred from line strengths gradients,
indicates that a metallicity spread is present in these systems.
This last argument applies to galaxies as a whole: whether or not their 
nuclei
host stellar populations with a spread in metal content depends on the
modalities of the galaxy formation. However, due to projection effects,
a substantial fraction of the light measured in the nuclei of Ellipticals comes
from regions located outside the three dimensional core. This
fraction can e.g. amount to $\sim$ 50 $\%$ for King models (Binney $\&$
Tremaine 1987). Therefore a pure radial metallicity gradient
translates into a metallicity spread in the stellar population
contributing to the light measured in the galactic nuclei.
\par
In this paper, the effect of a metallicity spread on
the integrated indices  is investigated, viewing a given galaxy 
(or a portion of it) as the  sum of SSPs. These models  are
compared to the relevant observations to derive conclusions on the stellar
content of the nuclear regions of early type galaxies and inferences on their 
formation process. Section 2 describes how the models are computed, and
the results are presented in Section 3. In Section 4 the 
models are compared to the observational data, and in Section 5 the
implications for the formation of elliptical galaxies are discussed. 
The main conclusions are sumarized in Section 6.

\section{Computational Procedure}
The spectral indices considered in the present work are defined as follows
(Burstein et al. 1984):
\begin{equation}
\Mgtw = - 2.5~{\rm Log}~\frac{F{_{\rm l}}(\Mgtw)}
{F{_{\rm c}}(\Mgtw)}  \label{eq:Mgd}
\end{equation}
\begin{equation}
\Hbeta = \Delta_{\beta} \times [1 -  
\frac{F{_{\rm l}}(\Hbeta)}{F{_{\rm c}}(\Hbeta)}]  \label{eq:Hbd}
\end{equation}
\begin{equation}
\Fetw = \Delta_{\rm Fe} \times [1 -  
\frac{F{_{\rm l}}(\Fetw)}{F{_{\rm c}}(\Fetw)}]  \label{eq:F2d}
\end{equation}
\begin{equation}
\Feth = \Delta_{\rm Fe} \times [1 -  
\frac{F{_{\rm l}}(\Feth)}{F{_{\rm c}}(\Feth)}]  \label{eq:F3d}
\end{equation}
where the various $F_{\rm l}$ denote the fluxes measured within the spectral 
windows of the different lines, centered at $\lambda \simeq$ 5175, 4863, 
5267 and 5334 \AA~ for the \Mgtw, \Hbeta, \Fetw and \Feth features, 
respectively. 
$F_{\rm c}$ are the pseudocontinuum fluxes measured at the line location, as 
interpolated from the fluxes measured in adjacent windows, and
$\Delta_{\beta}$, $\Delta_{\rm Fe}$ are the wavelength widths of the 
windows in which the 
\Hbeta ad Iron  indices are measured.\par
The above definitions apply to the spectra of single stars, of SSPs and of 
collections of SSPs, inserting the appropriate values for the fluxes. 
Therefore, for a collection of N simple stellar populations, 
each contributing a fraction $\Phi_{\rm S}$
to the total bolometric flux $F_{\rm bol}$ of the 
composite stellar population, the integrated indices are given by
equations (\ref{eq:Mgd}) to (\ref{eq:F3d}) with
\begin{equation}
F_{\rm l} = F_{\rm bol} \times \sum_{S=1}^{N}~(\frac{F_{\rm l}}
{F_{\rm bol}})_ {\rm S}~\Phi_{\rm S}      \label{eq:Fld}
\end{equation}
\begin{equation}
F_{\rm c} = F_{\rm bol} \times \sum_{S=1}^{N}~(\frac{F_{\rm c}}
{F_{\rm bol}})_{\rm S}~\Phi_{\rm S} \label{eq:Fcd}
\end{equation}
where the subscript S refers to the single SSP.
The spectral energy distribution of an SSP, and particularly the various
flux ratios, are controlled by a number of parameters, 
including the metallicity ($Z$), age ($t$), helium content ($Y$) and 
the elemental
abundances of the population. Thus, the integrated indices of a collection of
SSPs depend on how the fractionary bolometric flux 
$\Phi_{\rm S}$ is distributed over the range covered by all the relevant 
parameters. The problem of deriving these parameters from the observed
line strengths can be symplified considering various suitable indices,
each controlled by different parametes. For example, Gonzalez
(1993) uses a combination of Magnesium and Iron line strengths 
(mostly sensitive to $Z$) and the \Hbeta index
(mostly sensitive to $t$) to determine age 
and metallicity of a sample of early type galaxies. 
Still, in order to map the integrated indices of collections of SSPs
into their fundamental properties one needs to account for 
the presence of a possible spread in the parameters which control the
various line strengths. \par
One possible approach
to this problem consists in computing models for the chemical evolution of
galaxies, which automatically yield the distribution of the SSPs over the
fundamental parameters (e.g. Vazdekis et al. 1996, Tantalo et al. 1996). 
The output of these
models, though, depends on the specific ingredients used,
like the adopted star formation rate, initial mass function, ratio of
dark to luminous matter, nucleosynthesis, criteria for the establishment of 
a galactic wind, etc. A different approach consists in exploring
the dependence of the various indices on the presence of a spread in the 
populations parameters adopting a physically motivated function 
$\Phi_{\rm S}$, and using relations (\ref{eq:Mgd}) to (\ref{eq:F3d}). 
This approach, which has the advantage of allowing  
an easy exploration of the parameter space by simply changing the 
$\Phi_{\rm S}$ functions, will be adopted here.
Also, I will restrict to considering collections of SSPs all with  
the same age, but a substantial spread in metallicity. 
The results are then meant to describe the effects of the presence of
a metallicity spread in a stellar population formed within
a short time scale, so that the integrated indices are not 
appreciably influenced by age differences in the individual components.

\subsection{Line strengths for simple stellar populations}

In order to compute the integrated indices of composite stellar
populations one has to know the
$F_{\rm l}$/$F_{\rm bol}$ and $F_{\rm c}$/$F_{\rm bol}$ ratios 
of the SSPs as functions of $Z$ and $t$. 
Available SSP models in the literature tabulate bolometric corrections, colours
and spectral indices for different ages and metallicities. 
I then write:
\begin{equation}
\frac {F_{\rm l}}{F_{\rm bol}} = (\frac{F_{\rm l}}{F_{\rm c}}) \times 
(\frac{F_{\rm c}}{F_{\rm bol}})   \label{eq:Ide}
\end{equation}
for each SSP, and approximate $F_{\rm c}$ with the flux in the V band, for  
\Mgtw, \Fetw and \Feth, and with the B band flux for \Hbeta.
Although the pseudocontinuum fluxes do not correspond precisely to the flux in
the V or the B band, the main conclusions of this paper are not affected 
by this approximation. For example, the integrated \Mgtw indices computed
with $F_{\rm c}$ = $F_{\rm B}$ differ 
from those computed with $F_{\rm c}$ = $F_{\rm V}$ by
less than 2 percent.
\par
Two sets of SSP models are used here:
Buzzoni's models (Buzzoni et al. 1992; Buzzoni,
Mantegazza $\&$ Gariboldi, 1994; Buzzoni 1995a), 
and Worthey's models (Worthey 1994), hereinafter referred to as B and W
respectively. The metallicity range covered by
W models goes from 0.01\Zsun to $\sim$3\Zsun~
(\Zsun $\simeq$ 0.017), while B models encompass a larger $Z$
range, from $\sim$ 0.006\Zsun to $\sim$ 6\Zsun. Each metallicity  
is characterized by one value for the helium abundance: 
Worthey (1994) assumes $Y = 0.228 + 2.7 Z$, while in B models  
$Y$ increases less
steeply with $Z$, according to $Y \simeq 0.23 + Z$. Besides, the isochrones
used to construct the SSPs have solar abundance ratios. This corresponds to 
specifying the chemical trajectory followed in the evolution of the composite
stellar population. 
\par
The two sets of models present systematic differences in the bolometric 
output, broad band colours and line strengths (Worthey 1994,
Buzzoni 1995b). At least part of these differences can be ascribed to
the different choices of the $\Delta Y$/$\Delta Z$ parameter (Renzini 1995)
and to
the different fitting functions (i.e. the dependence of individual
stellar indices on effective temperature, gravity and metallicity) 
adopted in the computations. 
 
\subsection{The metallicity distribution functions}

The contribution of a given SSP to the bolometric
light of a composite stellar population of total mass $M_{\rm T}$ and total
luminosity $L_{\rm T}$ can be written as
\begin{equation}
\Phi_{\rm S} = \frac {L_{\rm S}}{L_{\rm T}} =  
(\frac {L}{M})_{\rm S}~\frac {M_{\rm S}}{M_{\rm T}}~
\frac {M_{\rm T}}{L_{\rm T}}.  \label{eq:Moverl}
\end{equation}
The distribution function $\Phi_{\rm S}$ is then proportional to the mass 
distribution over the metallicity range, through the inverse of the
$(M/L)_{\rm S}$ ratio, which depends on the metallicity and helium content 
(Renzini 1995).
\par
The closed box, simple model for the chemical evolution of galaxies
predicts that the distribution of the stellar mass over the total
metallicity follows the relation
$f~(Z) \propto e^{-Z/y}$ (Tinsley 1980), where {\it y} is the 
stellar yield.
According to this model, most stars are formed at the lowest 
metallicities. More sophisticated models for the chemical evolution of 
elliptical galaxies, which take into account the occurrence of galactic 
winds (Arimoto $\&$ Yoshii, 1987; Matteucci
$\&$ Tornamb\'e, 1987), also predict the existence of
a substantial metallicity spread in the stellar content of elliptical galaxies.
In these models, the more massive the galaxy, the later the galactic wind 
sets in, and further chemical enrichment is achieved. Correspondingly, 
as the galactic mass increases, the metallicity distributions get skewed 
towards higher $Z$ values. Nevertheless, a substantial fraction of low $Z$ 
stars is always present, and
indeed Arimoto $\&$ Yoshii's metallicity distributions (by number) 
for model
Ellipticals with mass ranging from 4 $\times$ 10$^9$ to 10$^{12}$ 
M$_{\odot}$ are well described by a closed box model relation :
\begin{equation}
f(Z) = \frac{{\rm exp}(-Z/y)}{\int_{Z_{\rm m}}^{Z_{\rm M}} 
{{\rm exp}(-Z/y)~dZ}}  \label{eq:Foz}
\end{equation}                   
with a minimum metallicity $Z_{\rm m} \simeq$ 0.01\Zsun, a maximum
metallicity $Z_{\rm M}$ increasing from $\sim$ 2\Zsun to $\sim$ 6\Zsun,  
and {\it y} varying from 2\Zsun to 3\Zsun. So, 
the metallicity range spanned by the stars
in these model ellipticals goes from $\sim$ 2.3 to $\sim 2.8$ dex.
\par
Relation (\ref{eq:Foz}) finds also observational support from the direct 
determination of the metallicity distribution of K-giants in the bulge
of our galaxy, which appear to follow a close box model relation with 
$Z_{\rm M}$ $\sim$ 5\Zsun and {\it y} $\simeq$ 2\Zsun (Rich 1988). McWilliam
and Rich (1994) revised Rich (1988) metallicity scale for [Fe/H] towards
values lower by $\sim$ 0.3 dex, but find a [Mg/Fe] overabundance
of the same amount, and state  that 
the distribution of [(Fe+Mg)/H] in the bulge stars may agree with
Rich (1988) [Fe/H] distribution.
\par
I then adopt eq. (\ref{eq:Foz}) to describing 
the $M_{\rm S}$/$M_{\rm T}$ distribution over the metallicity
for the composite stellar populations, and
explore the effect of different values for the three parameters 
$Z_{\rm M}$, $Z_{\rm m}$ and {\it y}. For a fixed (low) $Z_{\rm m}$, 
increasing values of
$Z_{\rm M}$ (and yield) are meant to describe stellar populations produced in 
an environment in which the chemical processing is terminated at progressively 
higher levels of completion. These sequences of models, then, conform to
the predictions of chemical
evolution models with galactic winds for increasing galactic mass.
Different values of $Z_{\rm m}$, instead, characterize different degrees of
pre-enrichment of the gas. 
\par
To derive the $\Phi_{\rm S}$ distribution the behaviour 
of the ($M/L$) ratio for the SSPs with increasing metallicity and helium 
content needs to be specified. In B models, which are characterized by a 
low $\Delta Y$/$\Delta Z$ parameter, $M/L$ increases with $Z$, going from
2.2 to 4.5 for $Z$ ranging from 0.01\Zsun to 3\Zsun, for the 
15 Gyr old
SSPs. In W 17 Gyr old models, the mass to (bolometric) luminosity ratio 
increases
mildly with metallicity up to a maximum value of 4.2 reached at
0.5\Zsun,and it decreases afterwards, down to 3.3 at 3\Zsun.
While the different behaviour reflects the different 
$\Delta Y$/$\Delta Z$ (Renzini 1995), these $M/L$ values are not directly 
comparable,
the total mass $M$ being computed with different prescriptions
(Worthey 1994). Besides this, neither of the two $M/L$ correspond
to what should be inserted in eq. (\ref{eq:Foz}) :
Buzzoni's $M$ values do not take into account the mass locked into
stellar remnants; Worthey's $M$ values do not include the
remnant masses for stars born with $M >$2 \Msun, but do include
both the remnant and the returned mass for stars born with mass between
the turn-off mass and 2\Msun. I then chose to
neglect the dependence of $(M/L)_{\rm S}$ on the metallicity of the SSPs,
and adopt $\Phi_{\rm S} \propto f(Z)$.
If $M/L$ increases with metallicity, as in B models, this approximation
leads to an overestimate of the contribution of the high metallicity SSPs
in the integrated indices for the composite stellar populations. If, on
the contrary, the $M/L$ given in W models are more appropriate,
the contribution of the high $Z$ populations will be underestimated. 
Notice, however, that in W models 
$M/L$ varies by only a factor of 1.2 for $Z$ varying from 0.01\Zsun to 3\Zsun.
In the following, composite stellar populations with this $\Phi_{\rm S}$
distribution function will be briefly referred to as CSPs.

\section {Results of the computations}
 
Following the prescriptions described in Sect. 2, the integrated
spectral indices are given by the following relations: 
\begin{equation}
\Mgtw = - 2.5~{\rm Log}~\frac {\int_{Z_{\rm m}}^{Z_{\rm M}}{dZ~~10^
{-0.4~({\rm Mg}_{2}^{\rm S}- BC_{\rm V}^{\rm S})}~e^{-Z/y}}}
{\int_{Z_{\rm m}}^{Z_{\rm M}}{dZ~~
10^{0.4~BC_{\rm V}^{S}}~e^{-Z/y}}}      \label{eq:mgcom}
\end{equation}
\begin{equation}
\Hbeta = \frac {\int_{Z_{\rm m}}^{Z_{\rm M}}{dZ~~\Hbeta^{\rm S}~10^
{0.4~BC_{\rm B}^{\rm S}}~e^{-Z/y}}}{\int_{Z_{\rm m}}^{Z_{\rm M}}{dZ~~
10^{0.4~BC_{\rm B}^{\rm S}}~e^{-Z/y}}}  \label{eq:Hbcom}
\end{equation}
\begin{equation}
\Fetw = \frac {\int_{Z_{\rm m}}^{Z_{\rm M}}{dZ~~\Fetw^{\rm S}~10^
{0.4~BC_{\rm V}^{\rm S}}~e^{-Z/y}}}{\int_{Z_{\rm m}}^{Z_{\rm M}}{dZ~~
10^{0.4~BC_{\rm V}^{\rm S}}~e^{-Z/y}}}         \label{eq:Ftwcom}
\end{equation}
\begin{equation}
\Feth = \frac {\int_{Z_{\rm m}}^{Z_{\rm M}}{dZ~~\Feth^{\rm S}~10^
{0.4~BC_{\rm V}^{\rm S}}~e^{-Z/y}}}{\int_{Z_{\rm m}}^{Z_{\rm M}}{dZ~~
10^{0.4~BC_{\rm V}^{\rm S}}~e^{-Z/y}}}         \label{eq:Fthcom}
\end{equation}
where $BC_{\rm B}$ and $BC_{\rm V}$ denote the bolometric corrections to the 
B and
the V band magnitudes, respectively, and the S index denotes the
SSP quantities, which depend on
metallicity and age. One can immediately notice that, since high $Z$ 
populations
yield a relatively low contribution in the B and V bands, the low metallicity
component of a composite stellar population will tend to dominate the
integrated indices. 
\par
The behaviour of these indices as functions of the average metallicity of
a composite stellar population has been explored by considering different
values for $Z_{\rm m}$ and $Z_{\rm M}$. 
In order to characterize each distribution in terms of one parameter,
an {\it average metallicity} has been computed, defined by the following
relation: 
\begin{equation}
[<{\rm Fe/H}>] = {\rm Log}~\frac {\int_{Z_{\rm m}}^{Z_{\rm M}}{dZ~~
\frac{(Z/X)^{\rm S}}{(Z/X)_{\odot}}
e^{-Z/y}}}{\int_{Z_{\rm m}}^{Z_{\rm M}}{dZ~~e^{-Z/y}}}.    \label{eq:fehc}
\end{equation}
Other definitions of {\it average metallicity} can be found in the
literature (see e.g. Arimoto $\&$ Yoshii 1987). In this respect, it should be
noticed that the quantity
$<[{\rm Fe/H}]>$ differs sistematically from [$<$Fe/H$>$] given by eq. 
(\ref{eq:fehc})
because the former assigns more weight to the low metallicity tail
of the distribution. The difference between the two quantities amounts to
$\sim$ 0.2 dex for the widest $Z$ distribution considered here ($0.01\Zsun <
Z < 6\Zsun$). The choice of using eq. (\ref{eq:fehc}) 
is motivated by the fact that this parameter better
describes the mass fraction of metals in the CSP, for the given 
$f(Z)$ distribution function. 
\begin{figure*}[tb]
\vspace {17cm}
\includegraphics{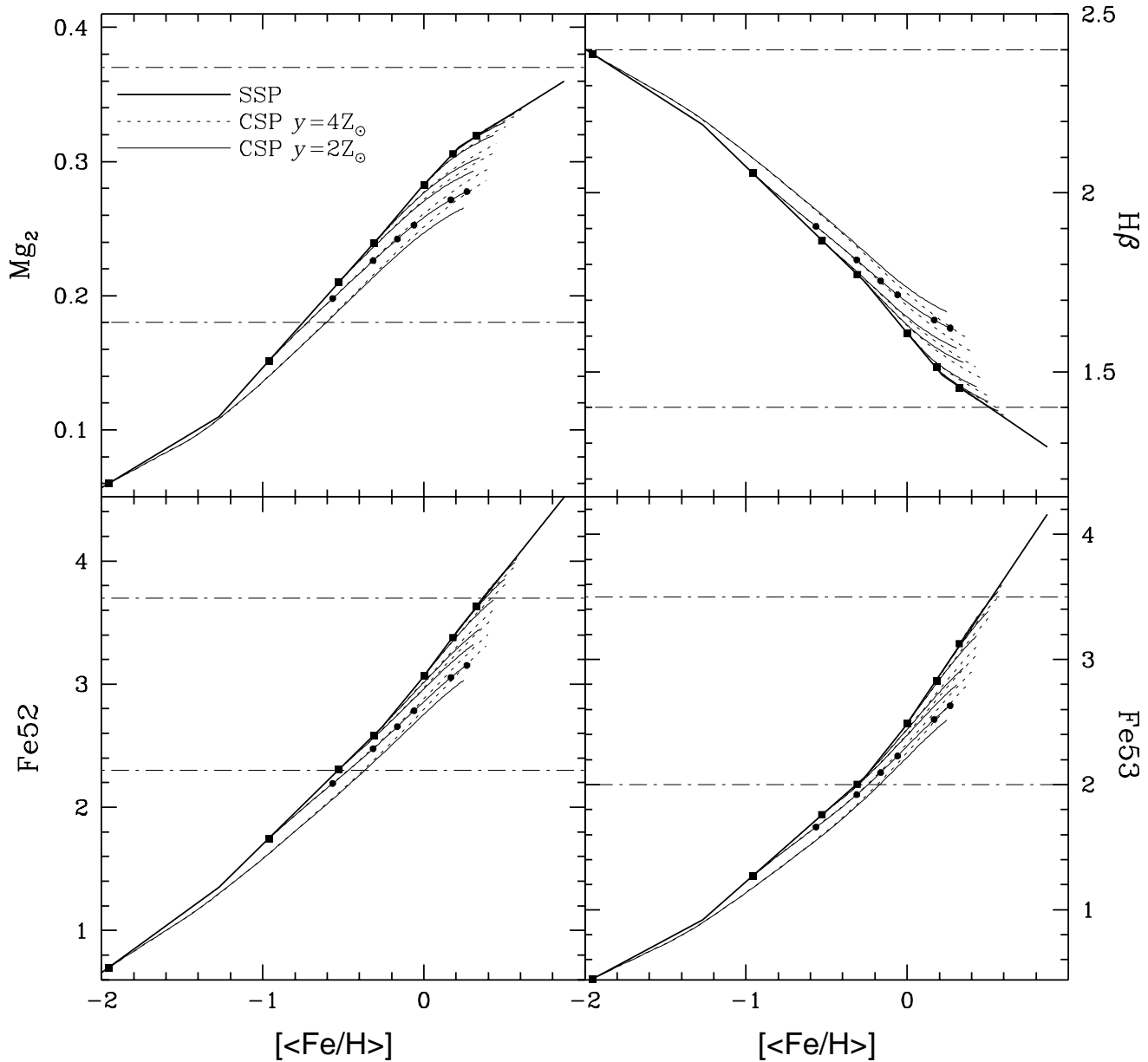}
\caption[]{Model line strenghts using Buzzoni's 15 Gyr old 
SSPs, as functions of metallicity. The thick lines are the loci of 
pure SSP models.
The thin lines display models for CSPs for various values of
the $Z_{\rm m}$ parameter. The following cases are shown:
$Z_{\rm m}$/\Zsun= (0.01,0.1,0.3,0.5,1,1.5,2), and the filled squares mark the 
values for SSPs with $Z = Z_{\rm m}$. 
Along each line, $Z_{\rm M}$ varies from $Z_{\rm m}$
up to 0.1.  The filled circles mark CSP  models with
$Z_{\rm m}$ = 0.1\Zsun, and $Z_{\rm M}$ = (0.5,1,1.5,2,4,6)$\times$\Zsun. 
The effect of different values of the parameter {\it y} is also shown. 
Finally, the dot-dashed lines show
the typical ranges spanned by observational data for the nuclei of elliptical
galaxies.}
\end{figure*}

\subsection {Indices versus metallicity using B models}

I will now describe the results of the integration of equations
(\ref{eq:mgcom}) to
(\ref{eq:Fthcom}) for the different $Z$ distributions, using B
and W sets of SSP models. Among the various models by Buzzoni
I have considered those computed with Salpeter IMF
and red horizontal branches. Analogous assumptions characterize Worthey's
models. 
\par
Figure 1 displays the integrated indices obtained using B SSPs at
15 Gyr, which are shown as thick lines in the four panels.
The thin lines illustrate the effect of the presence of a metallicity 
spread (see captions), the lowest of which corresponds to 
$Z_{\rm m}$ = 0.01\Zsun, and shows the
expected behaviour of the indices for galaxies of increasing mass, in the
frame of the wind models for the chemical evolution of galaxies. The 
different lines shows the results
obtained with different $Z_{\rm m}$, 
describing the effect of assuming different
degrees of preenrichment of the gas. 
\par
The loci described by the CSPs
for increasing average metallicity are shallower than the
pure SSP relations: this is due to the fact that the higher $Z$ populations
contribute less than the lower metallicity ones in the optical bands, where
the considered spectral indices are measured. As a consequence, the 
\Mgtw and Iron indices of these CSPs never reach 
values as high as those characteristic of the highest $Z$ SSPs, unless the 
metallicity spread is extremely small. 
Had I used the B-band flux as a measure of the
pseudocontinuum for calculating the integrated \Mgtw index the effect would
be stronger, since the high metallicity populations would receive even
less weight. 
Taking into account the $(M/L)_{\rm S}$ ratio dependence on $Z$ as given in B
models, when computing $\Phi_{\rm S}$, would also strengthen the 
difference between SSP and CSP models.
\par
At any given average metallicity, the \Mgtw and Iron indices for 
CSPs are weaker (and \Hbeta is stronger) than the corresponding 
values for SSPs. 
For example, a value of [$<$Fe/H$>$] = 0.22  can be obtained with a
composite stellar populations with parameters ($Z_{\rm m}$,$Z_{\rm M}$,$y$)=
(0.01,5,2)$\times$\Zsun. The differences between
the line strenghts of such a composite population and those of the SSP
with the same [Fe/H] amount to 
$\Delta$ \Mgtw $\simeq -$0.05, $\Delta$ \Hbeta $\simeq$ 0.18, 
$\Delta$ \Fetw $\simeq -$0.5 and $\Delta$ \Feth $\simeq -$0.4. These
results 
refer to the $\it y$ = 2$\times$\Zsun case. Adopting $\it y$ = 4$\times$\Zsun
(dotted lines), so as to enhance the fraction of the high $Z$ component, 
the differences are only slightly smaller. 
This effect is particularly important when dealing with
the highest metallicity galaxies.
\par
In much the same way, {\bf at any given value of the considered spectral 
index, the average metallicity of a composite stellar population
is higher than the metallicity of the SSP which has the same index}. 
This is illustrated in Figure 2, where I plot, as a function of \Mgtw,  
the difference ($\Delta$[Fe/H]) between the average metallicity of the 
CSPs and that of SSP models with the same value of the \Mgtw
index.
Along any line, the metallicity distributions have the same $Z_{\rm m}$
and increasing $Z_{\rm M}$, as in Figure 1. Figure 2 shows that 
the metallicity inferred from a given \Mgtw value using SSP models is
lower than what would be derived using CSP models, the difference being larger
the wider the metallicity distribution. The lower weight received by the high
$Z$ populations in eq. (\ref{eq:mgcom}) causes the rapid growth
of $\Delta$ [Fe/H] as the high \Mgtw ends of the curves are approached: 
high \Mgtw line strengths are obtained only with very large $Z_{\rm M}$ values.
This is not a small effect: at \Mgtw = 0.26 the difference in
the metallicities amounts to 0.3 dex, for a metallicity distribution
extending down to $Z_{\rm m}$ = 0.01\Zsun.

\begin{figure}[htb]
\vspace {9cm}
\includegraphics{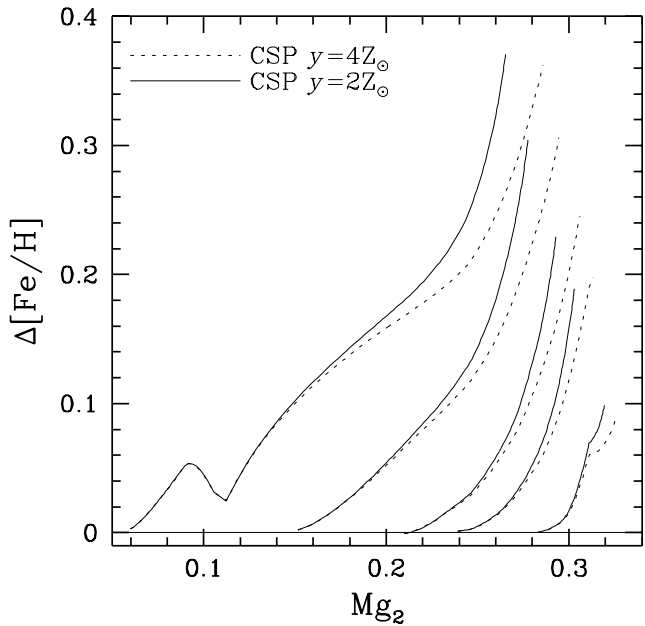}
\caption[]{The difference between the average metallicity of CSPs and the
metallicty of SSPs having the same \Mgtw index, as a function of the
index itself, for Buzzoni's 15 Gyr old models.
The different lines correspond to CSPs with different
values for the $Z_{\rm m}$ parameter: (0.01,0.1,0.3,0.5,1)$\times$\Zsun.
As in Fig.1, along the lines $Z_{\rm M}$ increases up to Z=0.1.
The minimum in the line corresponding to $Z_{\rm m}=0.01\Zsun$ 
is an artifact of the
linear interpolation between SSP models used in the computation.
The effect of adopting different {\it y} values is shown.}
\end{figure}

Adopting a large {\it y} the effect is still important,
while it gets small only reducing the widths of the metallicity distributions,
(i.e. increasing $Z_{\rm m}$), as obvious. It follows that
the calibration of \Mgtw in terms of metallicity via the comparison with
theoretical values for SSPs is affected by a systematic effect, which can
be considerably large, if a CSP spanning a 
substantial metallicity range is present. 
\par
The dot-dashed lines in Figure 1 show the typical range spanned by the 
observational data (e.g. Davies et al. 1987 ; WFG;
Carollo, Danziger $\&$ Buson 1993).
It appears that galaxies with (\Mgtw,\Fetw,\Feth) up to $\sim$ 
(0.26,3.,2.6) can be interpreted as hosting composite stellar populations with
a metallicity distribution as in the closed box model. For these
galaxies, higher metallic
line strengths correspond to larger values for $Z_{\rm M}$, and because
of  the Mg$-\sigma$ relation (Bender, Burstein $\&$ Faber 1993), to more 
massive objects. It should be noticed, however, that 
it is not possible to constrain the metallicity distribution from the 
integrated indices: the same value can be obtained with different metallicity
ranges for the  CSP, not for saying different distribution functions.
Nevertheless, it seems that
the observed metallic line strengths in the nuclei of  the less luminous 
Ellipticals are consistent with the theoretical expectations from
models for the chemical evolution which include the occurrence
of galactic winds.
\par
On the contrary, galaxy centers with metallic indices in excess than the above 
quoted values are not reproduced by these models (see also Casuso et al.
1996, Vazdekis et al. 1996), and appear instead to require some degree of 
pre-enrichment. The highest \Mgtw are
barely accounted for by pure SSP models, and the strongest Iron
indices are not reproduced by the CSP model with 
$0.1\Zsun \le Z \le 6\Zsun$ ,
in which the fraction of stars with sub$-$solar metallicity
is less than 0.4. As can be seen in Figure 1, adopting a yield as high as
4\Zsun does not alter these conclusions. 
\par
The \Hbeta values measured in the highest metallicity Ellipticals 
support this same picture: closed box models with 
$0.01\Zsun \le Z \le 6\Zsun$
have \Hbeta $\simeq$ 1.7, while galaxies with \Hbeta indices weaker than that
are observed. Besides, the models which account for the weakest metallic
indices, with $0.01\Zsun \le Z \le 0.5\Zsun$, are characterized by 
\Hbeta $\simeq$ 2, while galaxies with \Hbeta as large as 2.5 are
observed. Since \Hbeta is highly sensitive to age, one can interpret the 
data as an evidence for a younger age of the low metallicity 
Ellipticals (Gonzalez 1993). Notice however that the high $Z$ 
objects do require an old age. I will further discuss this point later.

\subsection {Indices versus metallicity using W models}

Figure 3 displays the models for composite stellar populations obtained using 
Worthey's set of SSP models, with an age of 12 Gyr. 
The legend is the same as in Figure 1. 
The highest metallicity considered in the $Z$ distributions
is now $\simeq$ 3\Zsun.
The qualitative effects of the presence of a metallicity
spread in the stellar population are the same as already discussed for
B models, and the indications derived from the comparison with the
observations are very similar. Also the quantitative effects are close
to those derived using for B models: the model with 
$0.01\Zsun \le Z \le 3\Zsun$ has (\Mgtw,\Fetw,\Feth,\Hbeta) $\simeq$
(0.23,2.8,2.5,1.7) for {\it y} = 2\Zsun, its average metallicity 
is [$<$Fe/H$>$] = 0.11, while a pure SSP with the same \Mgtw has
[Fe/H]=$-$0.16. Compared to Buzzoni's, W SSP models have
a steeper dependence of \Mgtw on [Fe/H] (see also WFG),
 higher \Feth and lower \Hbeta over all the metallicity range.
On the other hand for increasing [Fe/H], the bolometric corrections to the 
V and B bands decrease more rapidly in Worthey's models, leading to
a relatively lower contribution of the high $Z$ SSPs in 
equations (\ref{eq:mgcom}) to (\ref{eq:Fthcom}).
The two effects conspire to yield the same quantitative results on the 
integrated indices in the CSPs.

\begin{figure*}[tb]
\vspace {17cm}
\includegraphics{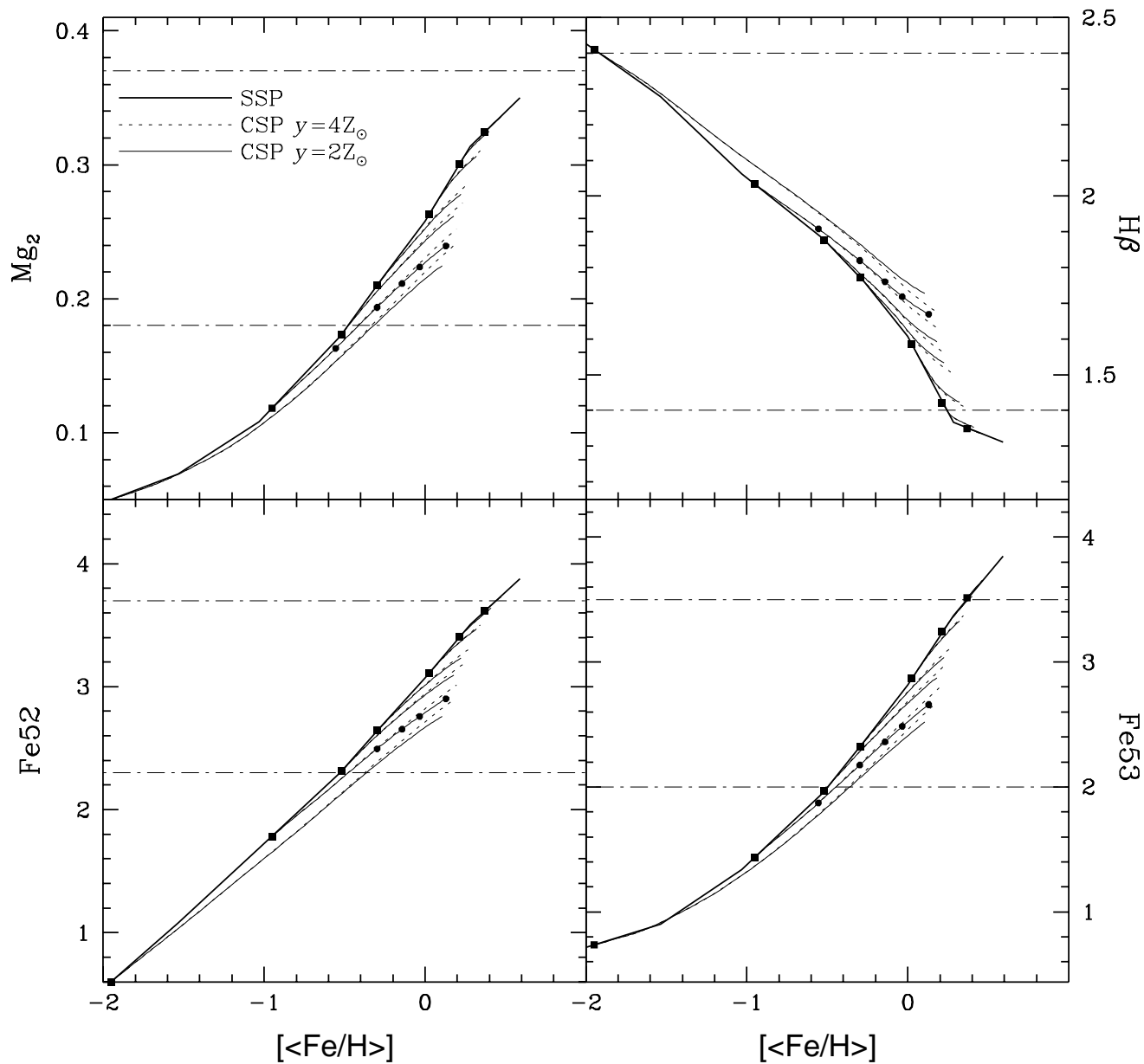}
\caption[]{The same as Figure 1, but for Worthey's 12 Gyr old SSPs.}
\end{figure*}

\section {Comparison of CSP models with the observations}

I will now compare more in detail the predictions of the CSP models 
illustrated in the previous section to the observations of the line strengths
in the nuclei of Ellipticals. I concentrate on some particular aspects
which can be crucial for understanding the modalities of galaxy formation. 
  
\subsection{The strong \Mgtw of Giant Ellipticals}

As shown in the previous subsection, a closed box model for the chemical
enrichment in which the gas is initially at $Z \simeq$ 0 fails to produce
a stellar population with \Mgtw as high as observed in the nuclei of
the brightest Ellipticals. One way to solve the
problem is assuming that the initial metallicity in the gas is larger
than 0. To investigate this in more detail, I have computed a sequence of 
models characterized by a fixed $Z_{\rm M}$ = 3\Zsun  and $Z_{\rm m}$ 
decreasing from $Z_{\rm M}$ to 0.05\Zsun. The results are shown in 
Figure 4 for W 17 Gyr and B 15 Gyr models, where
the hystogram shows the \Mgtw distribution of the Ellipticals in the
Davies et al.(1987) sample, which includes 469 objects, $\sim$ 93 $\%$ of 
which have \Mgtw in excess of 0.22. 

\begin{figure}[htb]
\vspace {10cm}
\includegraphics{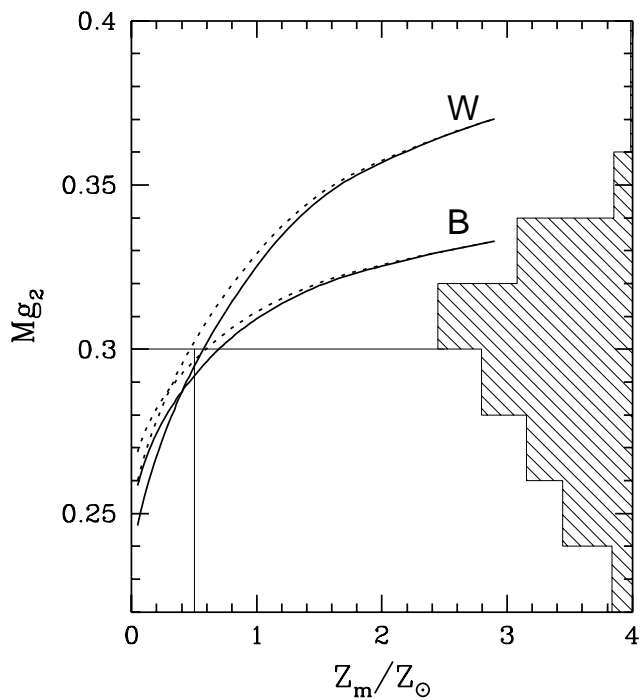}
\caption[]{Dependence of \Mgtw on the parameter $Z_{\rm m}$: the curves 
show the
loci of CSP models characterized by a maximum value of the metallicity
$Z_{\rm M}$ = 3\Zsun, while $Z_{\rm m}$ is made to decrease. Solid and 
dotted lines 
correspond to {\it y}/\Zsun = 2 and 3 respectively. The curves labelled 
W have been computed with Worthey's 17 Gyr old models; those labelled
B with Buzzoni's 15 Gyr old ones. Also plotted is the distribution of
the \Mgtw observed values for early type galaxies from Davies et al.
(1987). Objects with \Mgtw in excess of 0.3 require $Z_{\rm m}$ greater
than $\simeq$ 0.5\Zsun for both sets of SSP models.} 
\end{figure}

It appears that galaxies with \Mgtw higher than 0.3 (more than 40 $\%$ of
the total sample) require $Z_{\rm m}$ larger than 0.5\Zsun~for both W and
B models, in their nuclei. 
Integrated \Mgtw indices in excess of 0.32 are obtained 
with $Z_{\rm m}$ larger than $\sim$\Zsun, for W, or than 
$\sim$1.5\Zsun for B SSPs, and only the models computed
with Worthey's SSPs account for the largest \Mgtw. Notice that galaxies with 
nuclear \Mgtw as large as 0.4 
have been observed, and that the adoption of a higher value for the yield
(dotted lines) does not solve the problem. \par 
The difference in the results obtained using the two sets of models
is due to the different ages of the SSPs and to the 
different values of \Mgtw at the various metallicities.
Had I used Buzzoni's SSPs with $Z$ in excess of 3\Zsun, integrated \Mgtw 
larger than
0.32 would have required \Zsun $\le Z \le$ 6\Zsun. On the other hand,
using Worthey's 12 Gyr SSPs, the integrated \Mgtw for a metallicity 
distribution with 2\Zsun $\le Z \le$ 3\Zsun is $\simeq$0.33.
Therefore, these two sets of SSPs require that {\bf the nuclei of the strongest
\Mgtw ellipticals host both old and high metallicity SSPs, with a narrow
metallicity dispersion}. The older the population, the larger the allowed
metallicity range, but the presence of a significant component of
stars with $Z \le$\Zsun in galaxies with \Mgtw in excess than $\sim$ 0.32
seems unlikely, due to the strong effect it would have on the integrated
index.
\par
Weiss et al. (1995) derive metallic
line strengths for SSP models with total metallicity larger than solar, and 
different elemental ratios. For solar abundance ratios, their \Mgtw indices
happen to be in close agreement with those of B and W SSPs
at \Zsun, but are instead systematically larger at supersolar metallicities. 
For example their 15 Gyr SSP with $Z = 0.04$ has \Mgtw = 0.4, $\sim$ 0.07 dex
higher than B 15 Gyr model. Such high values would
relieve the problem of reproducing the data for the most 
luminous ellipticals. However, it is likely that a closed box metallicity 
distribution with $Z_{\rm m}$ $\simeq$ 0 would still predict too low 
\Mgtw values,
due to the higher contribution of the low $Z$ component to the optical light.
For example, enhancing by 0.1 dex the \Mgtw indices in W 17 Gyr old 
SSPs at $Z >$\Zsun, I obtain \Mgtw = 0.27 for a 17 Gyr old CSP with 
0.01\Zsun $\le Z \le$ 3\Zsun. The analogous
experiment with B 15 Gyr models yields \Mgtw = 0.30. Therefore, the need for
a small metallicity dispersion in the nuclei of the strongest \Mgtw Ellipticals
is a robust result. 

\begin{figure*}[tb]
\vspace {17cm}
\includegraphics{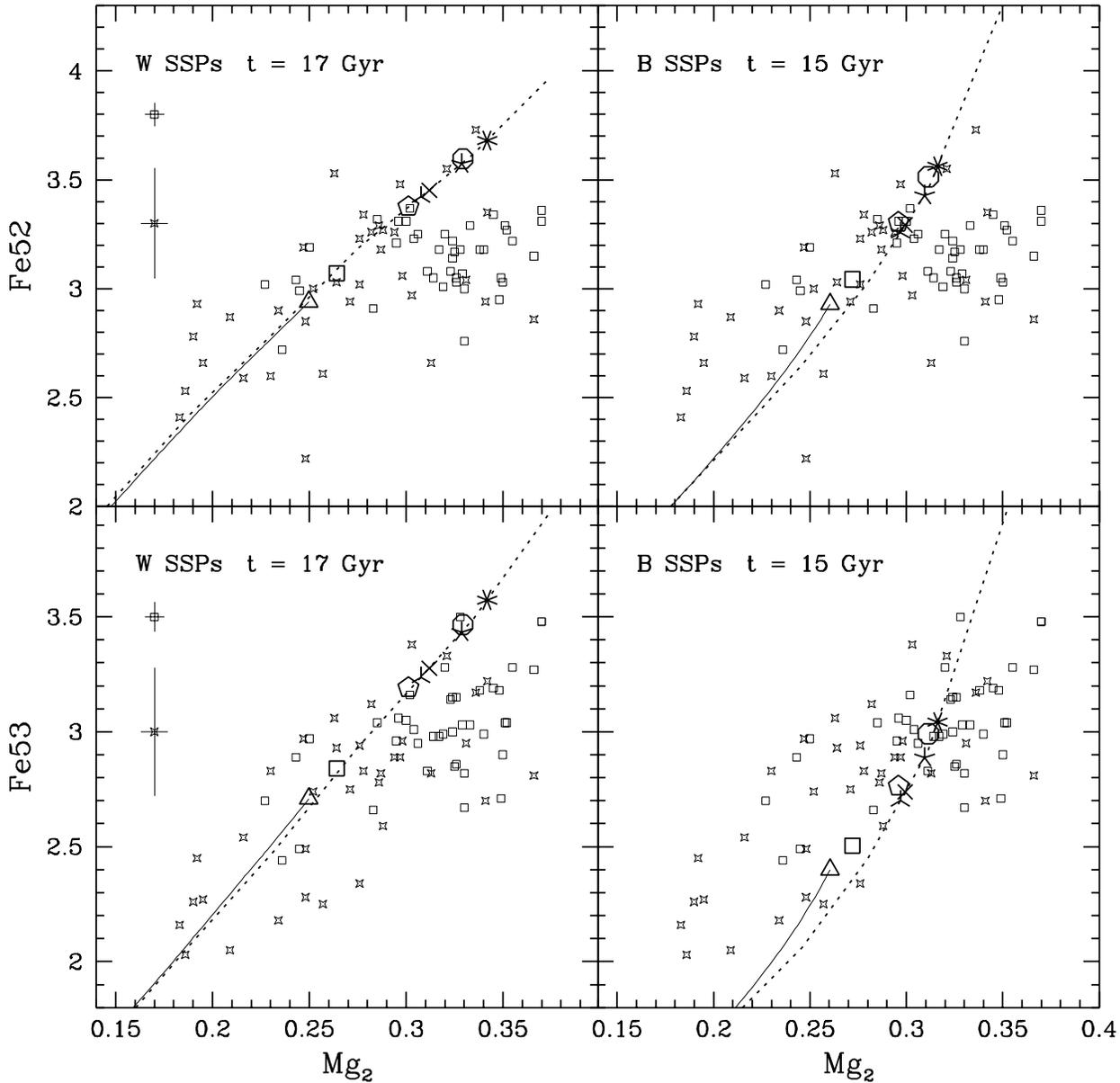}
\caption[]{Comparison between model predictions and observations of the
nuclear line strengths. The data points 
(shown as small symbols) are from Worthey, Faber $\&$ Gonzalez, 1992
(crosses) and Gonzalez, 1993 (squares). The error bars quoted by the
authors are shown in the two left panels, the smallest being relative
to Gonzalez data. 
SSP models by Worthey and Buzzoni are displayed as dotted lines. The large 
symbols show the line strengths for a selection of CSP models: 
$Z_{\rm m}$=0.01\Zsun (triangle), $Z_{\rm m}$=0.1\Zsun (square), 
$Z_{\rm m}$=0.5\Zsun 
(penthagon) and $Z_{\rm m}$=\Zsun (octagon), all having $Z_{\rm M}$ = 3\Zsun
and {\it y} = 3\Zsun. The sequence of CSP models obtained with 
$Z_{\rm m}$ = 0.01\Zsun and increasing $Z_{\rm M}$ is shown as a solid line.
Notice that the closed box simple model fails to account for the high
\Mgtw and Iron line strenghts shown by most of the data points.
The skeletal symbols show the indices of SSPs having [Fe/H]  
equal to the average metallicity of those CSP models shown as polygons with the
same number of vertices. For example, the cross shows the SSP line strengths 
for a metallicity equal to the average [Fe/H] of the CSP shown as a square.}
\end{figure*}

\subsection{Magnesium to Iron overabundance}

The plot of the iron line strengths as functions of \Mgtw allows to
check the internal consistency of the indications derived from the different
indices on the composition of the stellar population. 
In Figure 5, a selection of the models described in the previous 
section (large open symbols, see captions) is compared to the data 
from WFG and Gonzalez (1993)(small symbols), 
relative to the nuclear values of the indices. 
The solid lines show the locus described in this diagram from the
sequence of CSP models with $Z_{\rm m}$ = 0.01\Zsun and $Z_{\rm M}$ increasing
up to 3\Zsun. It can be seen that,
in these diagrams SSPs add as vectors: the effect
of a $Z$ distribution is that of shifting the model along the SSP line
(WFG) , to
an intermediate position between those corresponding to SSP models
with $Z = Z_{\rm m}$ and $Z = Z_{\rm M}$. 
Besides, the line strengths of SSP models are 
stronger than those of CSP models
with the same (average) metallicity: this can be seen in Figure 5
comparing the position of the skeletal symbols with that of the
corresponding polygons. Once again, this reflects the larger weight
of the low metallicity component on the CSP line strengths.
\par
It appears that the average location
of galaxies with \Mgtw$\le$0.3 is well reproduced by the models, better 
with Worthey's than with Buzzoni's SSPs. However, CSP models with  
$Z_{\rm m}$ = 0.01\Zsun barely reach \Mgtw $\sim$ 0.26, \Fetw $\sim$ 2.9, 
\Feth $\sim$ 2.7,
encompassing the range occupied by the weakest \Mgtw objects.
Galaxies with stronger metallic nuclear indices require some degree of
preenrichment in their centers, within this class of CSP models. 
Notice that this
is needed to account for both \Mgtw and Iron line strenghts.
\par
Those objects characterized by \Mgtw larger than approximately 0.3 (mostly
represented in Gonzalez's sample) depart from the general (\Fetw,\Feth) $-$
\Mgtw relation, exhibiting lower Iron indices than the model predictions, 
at the same magnesium index. This has been interpreted as evidence for a Mg/Fe
abundance ratio larger than solar in these (most luminous) ellipticals.
According to the current view of how the chemical enrichment proceeds in
galaxies, Magnesium is mostly produced by
massive stars exploding as Type II SNe, while a substantial fraction of
the Iron is provided by Type Ia SNe. Thus, a Mg to Fe overabundance can
be obtained either by increasing the relative number of Type II to Type Ia
events (e.g. with a flatter IMF), or by stopping the SF process
at early times, i.e. before a substantial amount of pollution by Type Ia 
events has taken place (WFG, Davies et al. 1993, Matteucci 1994).
Both scenarios predict that all the $\alpha$ elements, mainly produced in
massive stars, are overabundant with respect to Iron. It is then 
reasonable to assume that the [Mg/Fe] enhancement actually traces
and $\alpha$ elements overabundance (or an Fe underabundance), with
respect to the solar ratios. In the following section I will derive
a simple rule which enables to estimate the effect of an $\alpha$ elements
overabundance on the \Mgtw and Iron line strenghts.

\subsection{The effect of $\alpha$ element enhancement on the metallic
line strengths} 
 
The metallic line strength of SSPs are sensitive to the metallicity 
through two effects: one connected to the change of the overall shape of the 
isochrone, and the other to the dependence on the metal abundance of the 
specific feature in the individual stars which populate the isochrone,
described by the fitting functions. 
Both effects are such that enhancing the metal abundance, the metallic features
get stronger, but they operate in different ways. I try now to
estimate how the shape of the isochrone on one side, and the fitting 
functions on  the other, vary in response to an $\alpha$ elements
overabundance.
\par
The \Mgtw and \Fetw line strengths for SSPs are given by:
\begin{equation}
(\Mgtw)^{\rm S} = - 2.5~{\rm Log}~\frac{\int_{iso} {n(x)~{\it f}_{\rm c}(x)~
10^{-0.4~Mg_{2}(x)}
~dx}}{\int_{iso} {n(x)~{\it f}_{\rm c}(x)~dx}} \label{eq:mgssp}
\end{equation}
\begin{equation}
(\Fetw)^{\rm S} = \frac{\int_{iso} {n(x)~{\it f}_{\rm c}(x)~Fe52(x)~dx}}
{\int_{iso} {n(x)~{\it f}_{\rm c}(x)~dx}}  \label{eq:fessp}
\end{equation}
where $x$ describes the (Log g, Log T$_{\rm e}$) values along the isochrone,
$n(x)$ and $f_{\rm c}(x)$ are the number of stars and the contiunuum flux 
in the relevant wavelength bands, and $Mg_2(x)$, $Fe52(x)$ are the fitting 
functions, all quantities being computed at the point {\it x}.
\par 
For a given age, $n(x),f_{\rm c}(x)$ and the fitting functions depend on 
the total
metallicity $Z$ and on the fractions $\zeta_{\rm i}$ = $X_{\rm i}/Z$ of all the
different elements ($X_{\rm i}$ being the abundance by mass of the 
i-th element).
The effect of changing the fractions $\zeta_{\rm i}$ (at fixed total
metallicity $Z$) from solar ratios
to $\alpha$ enhanced ratios on the shape of the isochrone is likely to
be small. This has been shown to be valid (at low metallicities) for the
turn-off region by Chaboyer, Sarajedini $\&$ Demarque (1992)and 
Salaris, Chieffi $\&$ Straniero (1993). 
Considering the RGB, its location essentially depends on the abundance of
heavy elements with low ionization potentials (Renzini 1977), which,
being the main electron donors, control the H$^-$ dominated opacity.
The RGB temperature is then controlled by the total abundance of these
elements, namely Mg,Si,S,Ca and Fe (Salaris et al. 1993). 
The $\alpha$ enhanced mixtures predicted by detailed models for the
chemical evolution of galaxies (Matteucci 1992) are indeed characterized by
a similar value of the sum of the $\zeta_i$ fractions of
Mg,Si,S,Ca and Fe. For example, it varies from 0.2, at solar ratios,
to 0.18 for an $\alpha$ enhancement of $\sim$ 0.4 dex. This means that
a mixture with $Z=0.02$ and solar abundance ratios has the same
abundance (by mass) of electron donors as a mixture with 
$Z$ = 0.022 and an average enhancement of all the $\alpha$ elements of 
0.4 dex. It is then reasonable
to assume that the shape of the isochrone is mainly controlled by
the total metallicity, and that variations of the $\zeta_i$ fractions have
only a marginal effect, provided that all the $\alpha$ elements are enhanced.
Actually, compared to the solar ratios mixtures, the $\alpha$ enhanced 
ones in Matteucci (1992) models
have a slightly lower fraction of electron donors, implying slightly
warmer RGBs and, keeping fixed all the rest, lower metallic line
strenghts. The SSP models computed for solar ratios and 
$\alpha$ enhanced mixtures by Weiss et al. (1995) indeed 
support this conclusion (compare their model 7 to 7H).

\begin{figure}[htb]
\vspace {10cm}
\includegraphics{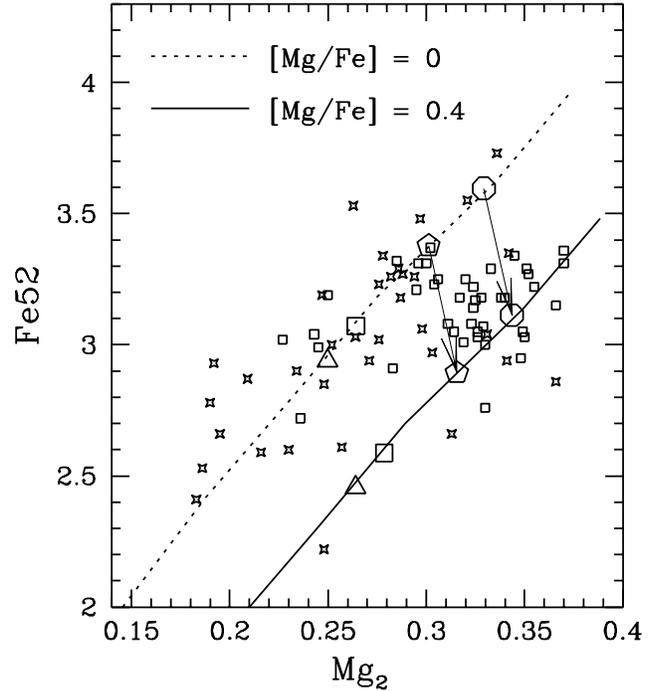}
\caption[]{The effect of an $\alpha$ elements enhancement on the \Mgtw and 
\Fetw line strenghts of W models with an age of 17 Gyr. 
The symbols are the same as in Fig. 5. 
The arrows show how the indices change for a progressively higher
[Mg/Fe], according to the simple scaling described in the text.
It appears that the highest \Mgtw galaxies require both a larger 
[Mg/Fe] ratio and a larger total metallicity.}
\end{figure}

I now turn to consider the metallicity dependence of the fitting
functions. Buzzoni's $Mg_2(x)$  and $Fe52(x)$ are expressed as the
sum of two terms, one depending only on metallicity and the other
on gravity and temperature. For such form, the \Mgtw and \Fetw indices for SSPs
can be written as:
\begin{equation}
(\Mgtw)^{\rm S} = \Theta_{\rm Mg}([{\rm Fe/H}]) + G_{\rm iso}  \label{eq:Mgs1}
\end{equation}
\begin{equation}
(\Fetw)^{\rm S} = \Theta_{\rm Fe}([{\rm Fe/H}]) + 
G^{\prime}_{\rm iso}  \label{eq:Fes1}
\end{equation}
in which $\Theta_{\rm Mg}$ and $\Theta_{\rm Fe}$ are exactly the 
dependences on [Fe/H]
of the $Mg_2(x)$ and $Fe52(x)$ fitting functions, while the $G$ functions
depend on the shape of the isochrone, and on how the stars distribute
along it. For B models, $\Theta_{Fe}$ = 1.15[Fe/H] and
$\Theta_{Mg}$ = 0.05[Fe/H] $\simeq$ 0.05[Mg/H], since Buzzoni's 
fitting functions have been 
constructed using a sample of stars with likely solar abundance ratios.
Asssuming that the $G$ functions only depend on the total metallicity $Z$, and
that the fitting functions only depend on the abundance of the element
contributing to the feature, for B models I can write:
\begin{equation}
(\Mgtw)^{\rm S} = 0.05~{\rm Log} \zeta_{\rm Mg} + h(Z)          \label{eq:Mgs2}
\end{equation}
\begin{equation}
(\Fetw)^{\rm S} = 1.15~{\rm Log} \zeta_{\rm Fe} + 
h^{\prime}(Z)   \label{eq:Fes2}
\end{equation}
were all the quantities dependent on total metallicity are described by
$h$, $h^{\prime}$, and $\zeta_{\rm Mg}$, $\zeta_{\rm Fe}$ are the 
contribution to the 
total $Z$ from Magnesium and Iron, respectively. 
It is now easy to derive the difference between the indices for two SSPs
with the same total metallicity (and age), but different $\alpha$ enhancements:
\begin{equation}
\Delta(\Mgtw) = a \times Log \zeta_{\rm Mg}^{(1)}/
\zeta_{\rm Mg}^{(2)}   \label{eq:Deltaam}
\end{equation}
\begin{equation} 
\Delta(\Fetw) = b \times Log \zeta_{\rm Fe}^{(1)}/
\zeta_{\rm Fe}^{(2)}   \label{eq:Deltaaf}
\end{equation}
with $a = 0.05$ and $b = 1.15$ for B models.\par
Gorgas et al. (1993) fitting functions, used in Worthey's models, do not 
allow the separation of the metallicity dependence from that on gravity
and effective temperature. Thus, equations (\ref{eq:Mgs1}) and (\ref{eq:Fes1})
are not strictly applicable. However, for bright RGB stars which dominate 
the \Mgtw index, Gorgas et al. fitting functions scale according to
$\Theta_{\rm Mg}$ $\sim$ 0.19[Fe/H] (WFG). 
Most of the contribution to the
\Fetw index comes instead from lower luminosity RGB stars (Buzzoni 1995a),
for which $\Theta_{\rm Fe}$ $\sim$ 1.5[Fe/H] seems a fair approximation.
Therefore eq. (\ref{eq:Deltaam}) and (\ref{eq:Deltaaf}) can be used to
estimate the effect of an $\alpha$ overabundance on W SSP models, with
$a = 0.19$ and $b = 1.5$, since solar elemental ratios are likely to
characterize also the stars in the Gorgas et al. sample.
\par
To summarize, the line strengths of SSPs with $\alpha$ elements enhancement
can be scaled from those of SSPs with the same age and total metallicity $Z$
(but solar elemental ratios) using relations (\ref{eq:Deltaam}),
(\ref{eq:Deltaaf}) provided that:\par\noindent
i) the shape of the isochrone (at given age and $Z$), and the 
distribution of stars along it, are the same;\par\noindent
ii) the dependence of the fitting functions on the metallicity is linear in 
[M/H] and can be separated from the other dependences;
\par\noindent
iii) a given index depends only on the abundance of the element which gives
rise to the considered feature. \par
These three requirements are never strictly true; nevertheless 
within the current understanding they seem to be valid approximations.
Chemical evolution models by Matteucci (1992) with an $\alpha$
elements enhanced mixture with [O/Fe] = 0.45 are characterized by
$\zeta_{\rm Mg}$/$\zeta_{\rm Mg,\odot}$ = 1.19 and 
$\zeta_{\rm Fe}$/$\zeta_{\rm Fe,\odot}$ = 0.48. It
follows that, for this kind of $\alpha$ enhancement, one expects that
at each metallicity point \Mgtw increases by $\sim$ 0.004 
and \Fetw decreases by $\sim$ 0.37 for B models. For W models
the expected differences are larger: \Mgtw increases by $\sim$ 0.015 
and \Fetw decreases by $\sim$ 0.48. 
The results of the application of these scaling relations are
shown in Figure 6, for W 17 Gyr old SSPs.
One can see that, as long as [Mg/Fe] is fixed, the theoretical indices 
describe a locus parallel to the solar ratio sequence. Therefore, the
flattening of the \Fetw vs \Mgtw relation at the high metallicity end
can be obtained only assuming different [Mg/Fe] ratios in the different
galaxies. The average (flat) slope of the
data points in the \Mgtw $>$ 0.3 domain requires an [Mg/Fe] increasing for 
increasing \Mgtw, up to an overabundance of $\sim$ 0.4 dex. 
It is worth noticing that, in this simple scheme, \Mgtw still remains
an indicator of total metallicity: changing the [Mg/Fe] ratio at constant 
$Z$ (i.e. enhancing $X_{\rm Mg}$ and decreasing $X_{\rm Fe}$ accordingly),
does not lead to a substantial increase of the \Mgtw index, while the effect on
the \Fetw index is more dramatic. As a result, the highest \Mgtw objects
are still accounted for by the highest total $Z$ populations.  
Finally, I notice that, in spite of leading to higher \Mgtw
indices for a given $Z$, an [Mg/Fe] = 0.4 is not sufficient to account for
the strongest \Mgtw values observed, unless the nuclei of these galaxies are
inhabited by virtually pure SSPs of large total metallicity. Supersolar
CSPs barely reach \Mgtw = 0.34, and metallicity distributions 
with $Z_{\rm m}$ = 0.5\Zsun don't go beyond \Mgtw = 0.32. Therefore, the need
for a small metallicity dispersion in the nuclei of the most luminous 
ellipticals still remains, even assuming an overabundance which accounts 
for their relatively low \Fetw indices.
\begin{figure*}[tb]
\vspace {11.5cm}
\includegraphics{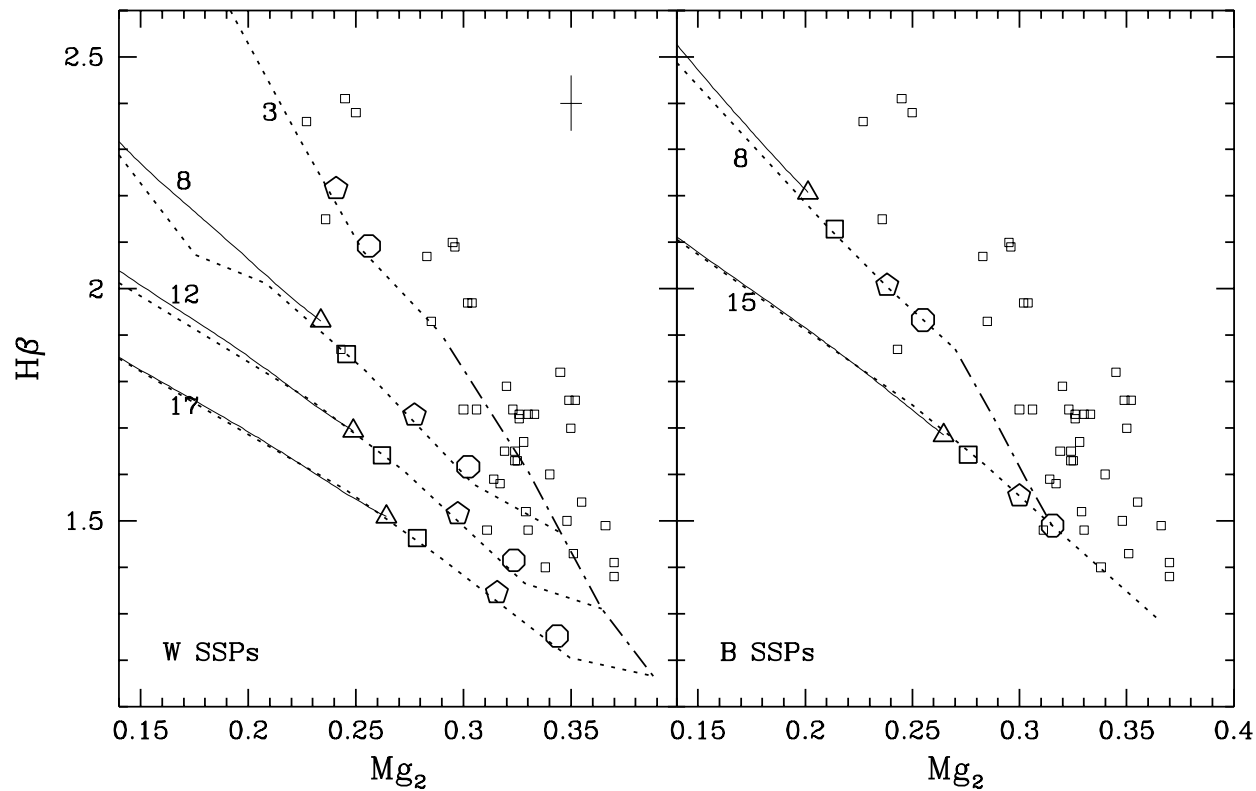}
\caption[]{Comparison between observations of
\Hbeta and \Mgtw indices in the nuclei of Ellipticals and model 
predictions using Worthey's (left panel) and Buzzoni's (right
panel) SSPs. The data points (small squares)
are from Gonzalez (1993) and the error bar quoted by the author is shown
in the left panel. The dotted lines connect SSP models of constant age
and different metallicities, each line labelled with its age in Gyr.
The dot-dashed lines connect constant metallicty SSPs with $Z$ = 3\Zsun
(left panel), 1.7\Zsun (right panel).
The solid lines are the loci described by single age CSPs, {\it y} =
3\Zsun, $Z_{\rm m}$ = 
0.01\Zsun, and $Z_{\rm M}$ increasing up to 3\Zsun, for the various ages.
The big symbols indicate the location in this diagram of CSPs with 
fixed Z$_{\rm M}$ = 3\Zsun and different Z$_{\rm m}$. 
The encoding is the same as 
in Figure 5, except for the following cases: the 3 Gyr old CSP shown as
a penthagon has $Z_{\rm m}$ = 0.6\Zsun (instead of 0.5\Zsun); the 
8 Gyr old CSPs based on Buzzoni's models have been computed with an upper
metallicity cut off of $Z_{\rm M}$ $\simeq$ 1.7\Zsun (instead of 3\Zsun).
This different limits are due to the limited Z range covered by the
SSP models at these ages. A constant shift of 0.015 (for
Worthey's models) and 0.004 (for Buzzoni's) has been applied to the
theoretical \Mgtw line strenghts to account for the [Mg/Fe] overabundance.}
\end{figure*}

\subsection{The \Hbeta index}

The \Hbeta line strength is very sensitive to the temperature of 
turn-off stars; thus plotting \Hbeta versus an index which 
is mainly controlled by $Z$ may allow to estimate indipendently age and
metallicity of a stellar system (see e.g. Gonzalez 1993). Adopting this 
approach, Faber et al. (1995) suggest that the nuclei of elliptical galaxies
form more an age sequence of high $Z$ objects,
as opposed to a metallicity sequence of old objects. 
In their comparison, the effect of a non 
zero [Mg/Fe] is taken into account plotting \Hbeta versus a newly 
defined index ([MgFe]), equal to the geometric mean of Mgb and $<$Fe$>$.
This index is meant to trace better the total metallicity $Z$, but there is no
guarantee that it actually does (see also Faber et al. 1995). I prefer to
use \Mgtw as metallicity indicator, and account for the Magnesium overabundance
with the simple scaling given by relation (\ref{eq:Deltaam}).
\par
Figure 7 shows the locus described by SSP and CSP models in the \Hbeta 
vs \Mgtw plane for various ages, together with Gonzalez (1993) data.
In order to mimic the effect of an $\alpha$ elements enhancement, a constant 
shift has been applied to the SSPs \Mgtw values, which amounts to 
0.015 dex for W and 0.004 dex for B models,
coresponding to a constant overabundance of [Mg/Fe] = 0.4. \par
The CSP models in Figure 7 appear to support Faber et al. (1995) conclusion,
and further show that a metallicity spread would substantially 
worsen the agreement between the models and the observations, 
at all metallicities. Actually, the dot dashed line in the
left panel of Figure 7, fairly fitting the observations, connects the 
$Z=$ 3\Zsun~SSPs in the W set of models. Since in this diagram SSPs
add as vectors, an internal age spread of $\sim$ 1-2 Gyr in the CSPs would 
hardly affect the interpretation of the data, which seem to require
younger average ages for the lower metallicity objects.
However, since the low \Mgtw
galaxies are also the fainter Ellipticals in the local sample, a tight
mass$-$age relation would be implied, with the less massive Ellipticals being 
(on the average) younger than the most massive ones. 
\par
Basically the same conclusion holds when considering B models 
(right panel): in spite of predicting \Hbeta line strengths which are
systematically higher than Worthey's, at any age and \Mgtw, still
the 15 Gyr old locus is too shallow with respect to the data. In essence,
for old ages the model \Hbeta index is too low at low metallicities, and 
ages younger than 8 Gyr are required to fit the \Hbeta values of the low
\Mgtw galaxies.
\par
The need for invoking an age difference in the galaxies of Gonzalez (1993)
sample is related to the mild dependence of the \Hbeta line strength on
the metallicity, which reflects the dependence on $Z$
of the turn$-$off temperature. This is the case for SSP models
with Red Horizontal Branches (RHB). On the other hand, as is well known,
the HB stars in the galactic globular clusters  become bluer
with decreasing metallicity, although the trend is not strictly
monothonic due to the {\it second parameter} problem (see e.g. Renzini 1977). 
If the average temperature of HB  stars increases for
decreasing metallicity, \Hbeta will be more sensitive to $Z$ than estimated
in the SSP models considered until now. As a result, the presence of a low 
metallicity tail in the stellar populations in Ellipticals
could affect the CSP \Hbeta line strenghts
appreciably, leading to higher equivalent widths the larger the
population in the low $Z$ tail. 
According to Buzzoni et al. (1994), an Intermediate Horizontal Branch (IHB)  
(corresponding to a temperature distribution peaked at Log T$_{\rm e}$ = 
3.82, and a blue tail extending up to Log T$_{\rm e}$ = 4.05, such as e.g. in
the globular cluster M3) leads to \Hbeta indices higher by
$\simeq$ 0.7 \AA, with respect to SSPs with RHB.
Thus, in order to estimate the impact of this effect on the \Hbeta
line strength I have computed integrated indices for CSPs
using Buzzoni's 15 Gyr old models with an artificially increased 
\Hbeta. The enhancement is taken equal to 0.5\AA~ for all metallicitites 
less than $\sim$ 0.6\Zsun, and linearly vanishing at \Zsun.
The effect of the adoption of an IHB on the \Mgtw index has instead been
neglected, since it is estimated to lead to a decrease of only a few
10$^{-3}$ mag (Buzzoni et al. 1994). Notice that a 
[Mg/Fe] = 0.4, which would correspond to an increase of \Mgtw by approximately
the same amount for B models, has not been taken into account in this 
computation. 
\par
The results are shown in Figure 8, where the locus described 
from CSPs with various $Z_{\rm m}$ is displayed. Obviously, 
those CSPs with
$Z_{\rm m}$ in excess of $\sim$ 0.5\Zsun are not affected by these new 
prescriptions, and their corresponding line indices lay on the
locus of RHB SSP models. On the contrary, CSPs with sufficiently
low $Z_{\rm m}$ have substantially higher \Hbeta, for a given \Mgtw,
and the data in this diagram could be interpreted as a sequence of
old stellar populations, with increasing average metallicity.
While the need for a small metallicity dispersion in the nuclei of
the most luminous Ellipticals discussed in the previous sections 
leaves little room for a sizeable contribution of a low $Z$
component, the less luminous Ellipticals can host a substantial low $Z$,
IHB component in their nuclei.  
This experiment shows that the strength of the \Hbeta
line is very sensitive to the temperature distribution assumed
for HB stars. Actually, just comparing Fig. 8 to the right panel
of Fig. 7, it appears that the data can be equally interpreted as
an age sequence, at a constant high metallicity,
or as a sequence at constant old age of CSPs with decreasing metallicity 
spread, corresponding to an increasing average metallicity. 

\begin{figure}[htb]
\vspace {11cm}
\includegraphics{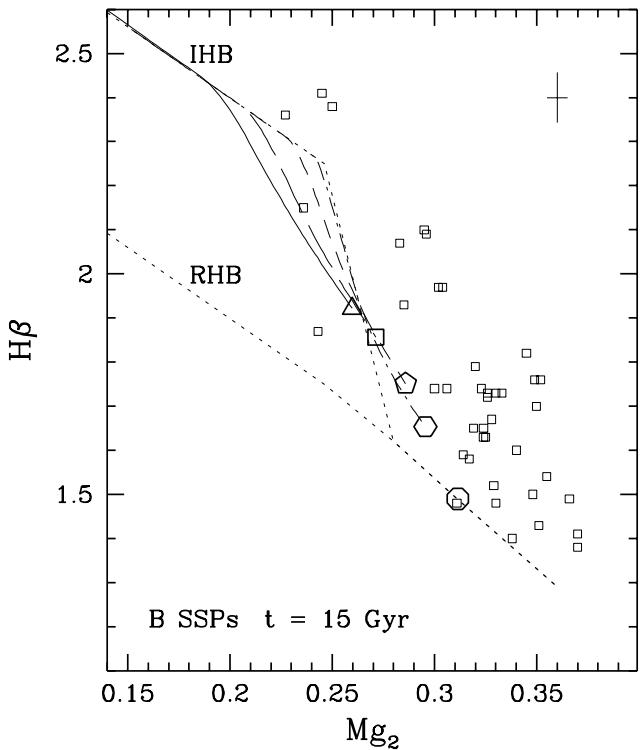}
\caption[]{Effect of the contribution of an IHB on the \Hbeta index. SSP models
from Buzzoni with an age of 15 Gyr are shown as dotted lines for a RHB
and an IHB (see text). CSP model sequences computed with IHB SSPs are shown
as different lines, along which $Z_{\rm m}$ increases up to 3\Zsun. 
All the CSP 
models have {\it y} = 3\Zsun, while $Z_{\rm m}$/\Zsun=0.01,0.1,0.3,0.5 
for the solid, long dashed, short dashed and dot$-$dashed lines, 
respectively. The last CSP model of each sequence is shown as a big symbol.
The big octagon, on the RHB line, shows the line strenghts for the
CSP model with $Z_{\rm m}$ =\Zsun, $Z_{\rm M}$ = 3\Zsun, {\it y} = 3.
No shift to higher \Mgtw values to
account for the possible [Mg/Fe] enhancement has
been applied here.}
\end{figure}

\section {Discussion}

The numerical experiments performed in this paper illustrate how the
\Mgtw, \Fetw, \Feth and \Hbeta line strengths are affected by the
presence of a metallicity distribution shaped like the closed box
model predictions. I have explored systematically the results of
changing the minimum and maximum metallicity ($Z_{\rm m}$, $Z_{\rm M}$) 
characterizing the chemical processing: the first parameter describes the
possibility of pre-enrichment of the gas in the closed box to different 
degrees; the second, the occurrence of galactic winds, inibiting further
chemical processing at various levels of completion.
I sumarize now the major results, and derive
some hints on the stellar population inhabiting the central regions of 
Ellipticals.
 
\subsection{The average metallicity in the nuclei of Ellipticals}

Due to its major contribution to the light in the optical bands, the
low metallicity component tends to dominate the integrated indices.
This implies that the metallic indices of the composite stellar populations
are systematically weaker than those of simple stellar populations,
with the same metallicity. Therefore, a given value for a metallic 
line strength, corresponds to CSPs with larger mass averaged metallicities
than SSPs. The difference between the two metallicities depends on
the width  of the $Z$ distribution in the CSP model, and it can be as
large as $\sim$ 0.3 dex.
As a consequence, a quantitative relation between
metallic line strength and average metallicity is subject to a substantial
uncertainty: it depends on the metallicity distribution, which
cannot be constrained, and on the SSP models, that may still be affected
by inadequacies. Indeed, some differences are present between the
various sets of models available in the literature (see also Charlot, Worthey
$\&$ Bressan 1996), which affect the relation between
integrated indices and the average metallicity of a given CSP.
This  impacts on the calibration of the spectral indices of different 
galaxies in term of
their metallicity, as well as on the derivation of abundance gradients
from line strength gradients within a given galaxy.
 
\subsection{The metallicity spread in the nuclei of Ellipticals}

The systematic exploration of the influence of the $Z_{\rm m}$ and 
$Z_{\rm M}$ parameters
shows that the sequence from low to high luminosity ellipticals (as far
as their central stellar population is concerned) can be 
interpreted in various ways: as a sequence of virtually pure SSPs of 
increasing metallicity; as a sequence of CSPs in which either the low 
metallicity component becomes less and less important, or both the minimum 
and the maximum metallicity increase. Chemical evolution models with 
galactic winds rather predict
a metallicity sequence among Ellipticals characterized by an increasing 
$Z_{\rm M}$ for increasing mass (and luminosity) of the
galaxy, with an important low metallicity component always present.
This class of models can account for objects with \Mgtw, \Fetw and \Feth up 
to $\sim$ 0.27, 3 and 2.7 respectively, while galaxies with higher values of 
these indices cannot be reproduced.
Notice that $\sim$ 80 $\%$ of objects in
the Davies et al.(1987) sample has \Mgtw $\ge$ 0.26. 
Since these considerations
apply only to the nuclear indices, classical wind
models can still be adequate to describe the global metallicity distribution
in Ellipticals, but a mechanism should be found to
segregate the high metallicity component in the nuclei.\par
A similar problem has been found by Bressan, Chiosi $\&$ Fagotto (1994),
when comparing the spectral energy distribution of their model Ellipticals 
with the observations. These authors pointed out that the
theoretical metallicity distribution was too heavily populated in
the low metallicity range, leading to an excess light between 2000 and
4000 \AA~with respect to the observed spectra. They noticed that a much 
better fit was achieved if a minimum metallicity of $Z=0.008$ was assumed,
and concluded that the classical chemical evolution models for 
elliptical galaxies, like those for the solar neighborhood, are affected by
the {\it G$-$dwaf problem}, i.e. the excess of low metallicity stars predicted
by the closed box model for the solar neighborhood. The classical 
solutions to {\it cure} the G$-$dwarf 
problem are (see Audouze $\&$ Tinsley 1976):\par\noindent
i) infall of metal free gas, so that the SFR exhibits a maximum
at some intermediate epoch, when a substantial enrichment has been already
accomplished;\par\noindent
ii) prompt initial enrichment (PIE), like what explored here by varying 
the $Z_{\rm m}$ parameter; \par\noindent
iii) adopting a SFR enhanced in high metallicity gas, in which
the stars are formed with a larger metallicity 
than the average $Z$ of the interstellar medium. \par
A variation of the PIE model consists in assuming that the first 
stellar generations are formed with a conveniently flat IMF (Vazdekis et al.
1996), so that they 
contribute metals at early times, but not light at the present epoch. 
The following
generations would instead form with a normal IMF.\par
All these are in principle viable solutions, and which, if any, of these
applies to the nuclei of Ellipticals remains to be seen. However, I
notice that the infall models predict for the most massive galaxies 
\Mgtw indices not larger than 0.28 (Bressan et al. 1996): 
still a low value with respect to the
observations in the nuclei of massive Ellipticals.
Solution iii) requires large inhomogeneities in the gas, and a variable
IMF seems a rather {\it ad hoc} solution. \par
A prompt initial enrichment 
for the gas in the nuclei of Ellipticals is easily realized by relaxing the
hypothesis of istantaneous complete mixing of the gas, and allowing
enriched gas to sink to the center. This is indeed a natural result of
dissipational galaxy formation (Larson 1975). 
During its formation process, a galaxy consists
of two components: one dissipationless (the newly formed stars), and
one dissipative (the gas). Once formed, the
stars stop participating to the general collapse, keeping thereafter
memory of their energy state at formation. The gas, instead, will
continue to flow towards the center, being progressively enriched as
SN explosions take place. Thus, gas accumulated in the nuclear regions is
pre-enrinched by stars which formed (and died) in the outer regions, and
will further form stars, which will then be the most metal rich in the
galaxy. The metal poor stars missing in the galaxies nuclei should be
found in the outer regions. In other words, the different dissipation 
propertities of the stars
and the gas would lead to a chemical separation within the galaxy, during its
formation, no matter wether the protogalaxy is a monolithically collapsing
cloud, or if it consists of gas rich merging lumps. Interestingly, 
this out$-$inward formation, would also allow for 
peculiarities in the core kinematics, as observed in a substantial fraction 
of galaxies (Bender et al. 1993), since the formation of the nucleus
would be partially decoupled from the formation of the rest of the galaxy. 

\subsection{The elemental ratios in the nuclei of Ellipticals}

The Iron indices vs \Mgtw plot suggests the presence of a Magnesium 
overabundance in the brightest Ellipticals (\Mgtw $>$ 0.3), 
as pointed out by WFG comparing the data to SSP models. The presence 
of a metallicity distribution does not alter this conclusion, since the CSP 
models considered here
describe the same locus of SSPs in the Iron vs \Mgtw diagram. If the
Magnesium overabundance is tracing a true $\alpha$ elements overabundance,
it is possible to estimate the enhancement quantitatively using
a simple scaling of the SSP models constructed for solar abundance ratios.
This estimate however is very sensitive to the dependence of the \Mgtw index
on the Mg abundance, and of the Iron index on the Fe abundance.
Assuming that these dependences are the same as the metallicity dependence 
of the relative fitting functions, for Worthey's models I have found that
the brightest ellipticals should be characterized by [Mg/Fe] ratios
ranging from 0 to 0.4 approximately, for increasing \Mgtw. It may
seem that a Magnesium overabundance would easily explain the strong \Mgtw 
values
without invoking high total metallicities in the nuclei of the brightest
Ellipticals. However, [Mg/Fe]$>$0 means Mg enhancement together with Fe
depletion, with respect to the solar ratio. In the frame of an overall
$\alpha$ element overabundance, the chemical evolution
models actually predict a lower fraction of electron donors at fixed
total metallicity, as mentioned in Section 3.4. 
Correspondingly, the temperature of the RGB is increased,
counterbalancing the increase of the \Mgtw index conveyed by a higher
Mg abundance. Thus, large metallicities and small metallicity dispersions 
are still needed to account for the data.\par
According to current modelling an overabundance of
[Mg/Fe]=0.4 implies very short formation
timescales for the whole galaxy, since the Mg and Fe gradients, within the
errors, are the same (Fisher, Franx $\&$ Illingworth 1995). 
How short is interesting to look at: 
Matteucci (1994) models
for the chemical evolution of Ellipticals predict that a solar [Mg/Fe] is 
reached already at 0.3 Gyr. An overabundance implies formation timescales 
shorter
than this, and the higher [Mg/Fe], the shorter the timescale required.
Indeed, in Matteucci (1994) {\it inverse wind models}, the galactic
wind occurs at only $\sim$ 0.15 Gyr in a 5 $\times$ 10$^{12}$ M$_\odot$ galaxy,
and yet the corresponding overabundance is not larger than 0.3. The formation
timescales inferred from a given [Mg/Fe] overabundance depend on the
adopted model for the Type Ia supernovae progenitors, and accordingly can
be considered quite uncertain. It seems however unlikely that a Mg to Fe
overabundance could be accomplished with formation time scales longer
than $\sim$ 1 Gyr. For example, Greggio (1996) finds that, following a 
burst of Star Formation, $\simeq$ 50$\%$  of the total Iron from the 
Type Ia SNe is released  within a timescale ranging from 0.3 to 0.8 Gyr,
for a variety of SNIa possible progenitors.

\subsection{The \Hbeta line strength as age indicator}

As for the Iron indices, the comparison of models to observations in the 
\Hbeta vs \Mgtw plot
does not change if a metallicity spread in the stellar populations is
taken into account. In SSPs with red horizontal
branches the only way to enhance \Hbeta is by adopting a warmer turn$-$off,
that is younger ages. In this case, the constraint from the metallic
indices on the metallicity and metallicity dispersion discussed until now
is even stronger, since younger ages make \Mgtw, \Fetw and \Feth weaker.
Extremely large metallicities should characterize the nuclei of all
Ellipticals in Gonzalez (1993) sample, with virtually no 
metallicity dispersion.
If however the temperature distribution of
HB stars is wide enough, strong \Hbeta indices can be obtained in
old stellar populations, due to the contribution of {\it warm} HB stars. In
this case, the \Hbeta vs \Mgtw plot would trace the different proportions
of this stellar component. On the average, the effect is stronger 
for the less metallic galaxies, if they host a composite
stellar population with larger fraction of low $Z$ stars, and the data are 
consistent with an old age for the stars in the
centers of this sample of Ellipticals \par
The nuclei of
galaxies with \Mgtw in excess of $\sim$ 0.3 are however likely to host
only stars with metallicity larger than $\sim$ 0.5\Zsun. 
For these galaxies, there is little room for a warm
HB component, at least in the frame of the canonical stellar evolution,
and yet their \Hbeta line strengths range from $\sim$ 1.4 to $\sim$ 1.8 \AA.
Adopting Worthey's 3\Zsun~SSP models, this implies and age range from
$\sim$ 12 to $\sim$ 5 Gyr, if the stellar populations in the nuclei of
these galaxies are coeval. If, however, I consider a two component population,
one very old and one very young, relatively high \Hbeta values can be
accomplished with a small contribution to the total light (and even smaller
to the total mass) from the young 
component. For example, a combination of a 17 Gyr plus
a 1.5 Gyr old SSP, both with 3 times solar metallicity, contributing 80
$\%$ and 20 $\%$ of the light respectively, has \Hbeta $\simeq$ 1.74 and
\Mgtw $\simeq$ 0.32. The contribution to the total mass of the young SSP
would amount to only $\sim$ 7 $\%$. 
Adopting a solar metallicity for the young component,
one gets \Hbeta $\simeq$ 1.8, \Mgtw $\simeq$ 0.32 with only 10$\%$ of the
light coming from the 1.5 Gyr old population, corresponding to a
$\sim$ 3$\%$  contribution to the total mass. Indeed, owing to the steep age 
dependence of \Hbeta, a small fraction of light from the young stellar
population is sufficient to enhance the \Hbeta line strengh.
If this was the case, the bulk of the population in the nuclei of
these galaxies would be truly old, with \Hbeta tracing a relatively
unconspicuous recent Star Formation event. 

\section {Conclusions}

Using the SSP models currently available in the literature to construct 
integrated indices for composite stellar populations with a metallicity
spread I have shown that the nuclei of the most luminous elliptical galaxies
should host stellar populations with:
\par\noindent
a) high total metallicity; 
\par\noindent
b) a Magnesium overabundance with respect to Iron, with varying 
degrees of the [Mg/Fe] ratio. \par\noindent 
c) a small metallicity spread.\par
Condition a) is met by processing the gas through multiple
stellar generations, and condition b) requires that this processing occurs
within a short time scale. This inevitably means that during the
chemical enrichment the star formation rate was very high, implying
a correspondingly large SNII rate. In order to proceed with the chemical 
processing, the gas had to be subject to confinement, that is it had to be 
located within a deep potential well. 
Condition c ) is met if the gas turning into stars in the nuclei of 
Ellipticals has been substantially pre$-$enriched, or if the maximum SFR was 
achieved at some
late stage, when partial chemical processing had already been completed, like
in infall models. However, for any behaviour of the SFR with time, 
as long as one considers a selfpolluting gas mass, a low metallicity 
component in the final stellar population is unavoidable: it is this
low metallicity component which provides the metals to build up the
high $Z$ stars. As a consequence, the extremely high \Mgtw indices would
rather favour the pre$-$enrichment alternative.
\par
These facts support the notion that 
the gas out of which the nuclei of the most luminous ellipticals
formed was produced within the galaxy itself, and was not accreted from the
outside. 
Several evidences indicate that merging must have played a role in the
formation of these galaxies, including the relatively shallow metal
line gradients (e.g. Davies et al. 1993), and the 
peculiar kinematics in the
nuclei of a substantial fraction of galaxies (see Bender 1996). 
The indications from the analysis performed in this paper
suggest that the merging subunits should have been mostly gaseous, and
confined within the deep potential well in which the galaxy itself is 
found. 
The segregation of the high $Z$ component in the inner parts
could result from the gas partecipating of the general collapse more than
the stars, which could have formed within the merging subunits.
\par
As discussed in the previous section, the indications from the \Hbeta line 
strength are sufficiently ambiguous that the possibility that the bulk
of the stars in the nuclei of the brightest Ellipticals are indeed old 
remains favourable.
Therefore, the formation of these galaxies should 
have occurred at high redshifts, both the stellar component and the potential
wells, within which the chemical processing could proceed up to
high metallicities in short time scales. 
\par
For the lower luminosity Ellipticals the conclusions are more ambiguous,
as their nuclear line strengths are consistent with both wide and
narrow metallicity distributions. However, galaxies with central \Mgtw
in excess of $\sim$ 0.27 (approximately 70$\%$ in Davies et al. sample)
are not accounted for by closed box models with $Z_{\rm m} \sim$ 0.
Therefore, for the majority of Ellipticals some chemical separation 
should have taken place during their formation, although a substantial
low metallicity component could be present in their nuclei. If this was
the case, the \Hbeta
line strength would trace the proportion IHB stars produced by this 
component of the composite stellar population in the different
galaxies, which would then be old. Finally, the Mg/Fe ratio in the
lower luminosity Ellipticals
is likely to be solar, suggesting longer timescales for 
the formation of the bulk of their stars, with respect to
the brigther Ellipticals. 
\par
A final caveat concernes the reliability of the SSP models used for
the interpretation of the observational data. The differences among the 
various 
sets of models, the different fitting functions,
and the lack in the stellar data sets of a fair coverage
of the age, metallicity and [Mg/Fe] ratio cast some doubt on the
use of these models to derive quantitative informations. 
The data seem
to suggest that the stars in the nuclei of elliptical galaxies have
\Mgtw indices stronger than what can be reasonably inferred using these
SSP models. Actually \Mgtw as large as 0.4
are difficult to account even with the oldest and most metal rich SSPs
in the Worthey's and Buzzoni's sets of models. Although this may be a 
problem of just a few object, this fact may suggest some inadequacies in  the
SSP models. If the dependence of the \Mgtw index on metallicity is currently
underestimated at the high $Z$ end,
lower total metallicities, larger metallicity
dispersions and lower [Mg/Fe] ratios would be possibly allowed for the
stellar populations in the nuclei of the brightest ellipticals. Nevertheless,
the presence of a small metallicity dispersion in the nuclei of giant
ellipticals and the need for old ages seem to be quite robust conclusions,
due to the larger contribution to the optical light of the less metallic
populations.
 

\section {Acknowledgements}

It's a pleasure to thank the whole staff at the Universitaets Sternwarte$-$
Muenchen for the kind and generous hospitality, and in particular the
extragalactic group, who made my research work specially
enjoyable for the cheerful environment and stimulating scientific discussions.
I am particularly grateful to Ralf Bender and Alvio Renzini for many
enligthning discussions on this work and careful reading of the manuscript.
The Alexander von Humboldt--Stiftung is acknowledged for support.

\end{document}